\documentclass[jou,12pt,floatsintext]{apa7}

\usepackage[american]{babel}

\usepackage{csquotes} 
\usepackage[style=apa,sortcites=true,sorting=nyt,backend=biber]{biblatex}
\DeclareLanguageMapping{american}{american-apa} 
\usepackage[T1]{fontenc} 
\usepackage{amsmath, amsthm}
\usepackage{dsfont}
\usepackage{interval}
\usepackage{mathptmx} 

\newtheorem{assump}{Assumption}

\newtheorem{prop}{Property}

\newcommand{\ind}{\perp\!\!\!\!\perp} 

\title{The positivity assumption in causal mediation analyses? Checked!} 
\shorttitle{Mediational positivity verification} 
\leftheader{Chatton et al.}
\authorsnames[{1,2,3,4},5,{6,3,7},{1,2}]{Arthur Chatton, Geneviève Lefebvre, 
Mireille E. Schnitzer, Denis Talbot}
\authorsaffiliations{{Département de Médecine Sociale et Préventive, Université Laval, Québec, QC, Canada}, {Centre de Recherche du CHU de Québec - Université Laval, Québec, QC, Canada}, {Département de Médecine Sociale et Préventive, Université de Montréal, Montréal, QC, Canada}, {Centre de recherche Azrieli, CHU Sainte-Justine, Montréal, QC, Canada}, 
{Département de Mathématiques, Université du Québec à Montréal, Montréal, QC, Canada},
{Faculté de Pharmacie, Université de Montréal,  Montréal, QC, Canada},
{Department of Epidemiology, Biostatistics and Occupational Health, McGill University, Montreal, QC, Canada}}
\date{\today} 
\course{PSY 4321} 
\professor{Dr. Professor}  
\authornote{\addORCIDlink{Arthur Chatton}{0000-0002-0018-5899} was initially supported by a CRM-StatLab-CANSSI postdoctoral fellowship.
Geneviève Lefebvre is an FRQS Senior Research Scholar.
Jesse Gervais is supported by a doctoral scholarship from the FRQNT.
Mireille E. Schnitzer is a Canada Research Chair (Tier 2) in Causal inference and Machine learning in Health Sciences.
\addORCIDlink{Denis Talbot}{0000-0003-0431-3314} is an FRQS Junior-2 Research Scholar. 
We have no conflicts of interest to disclose.
The dataset used for the tutorial is available at icpsr.umich.edu/web/ICPSR/studies/29583/versions/V1. This dataset has been used in previous studies by numerous research groups.
Contact: \addORCIDlink{Arthur Chatton}{0000-0002-0018-5899}, École de Santé Publique de l'Université de Montréal, 7101 Av. du Parc, Montréal, QC H3N-1X9, Canada. email: arthur.chatton@umontreal.ca}

\abstract{Causal mediation analyses are increasingly used in psychological sciences. Among the required assumptions, positivity is unfortunately seldom mentioned, likely due to the lack of tools for checking it. Mediational positivity is more complex than positivity in standard, non-mediated exposure effect analysis, because it requires positivity for both the exposure and the mediator, and because the specific form of the positivity assumption depends on the mediation estimand of interest. We propose an extension of the Positivity Regression Trees (PoRT) algorithm — which was recently designed to check positivity in non-mediated settings without requiring assumptions about the modeling or the data-generating process — to controlled, natural and interventional mediational effects. We illustrate its use through an application in mental health and have made it accessible through the \texttt{port} \texttt{R} package and a related notebook available at github.com/ArthurChatton/dePoRT-notebook. Finally, we discuss consequences and provide recommendations for when positivity violations are identified.}

\keywords{Assumption, Causal inference, Data support, Decision tree, Mediation analysis.} 

\begin{document}
\maketitle 


The rise of the potential outcome framework  \parencite{Neyman_1923, Rubin_1974} has profoundly changed the way we study causal questions. From the definition of causal estimands (i.e., what is the question?) to the non-statistical assumptions needed to identify causal estimands (i.e., can we answer the question?), this framework has led to improvements in how we think about causality, how we teach it, and how we estimate causal effects \parencite{Lundberg_2021}. 

Mediation analysis, the study of causes' mechanisms \parencite{Wright_1934, Judd_Kenny_1981, Baron_Kenny_1986}, has also benefited from this causal revolution.\footnote{The link between traditional and causal mediation analyses is extensively discussed in \textcite{MacKinnon_2020}.} In a seminal paper, \textcite{Robins_Greenland_1992} were the first to use the potential outcome notation $Y^{a,m}$ to represent the outcome $Y$ that would have been observed if the exposure $A$ been set to some value $a$ and the mediator $M$ been set to $m$, yielding major developments in our understanding of the assumptions needed to identify mediational effects \parencite[reviewed in][]{Nguyen_2021}. Among these assumptions, exchangeability (or no unmeasured/residual confounding) has received a lot of attention \parencite[e.g.,][]{Vanderweele_2015}, while positivity has been left far behind. 

Positivity violations occur for two reasons \parencite{Petersen_2012}. There may be a structural reason why no individual experiences the exposure (i.e., practical violations), or a particular subgroup may lack exposures by chance in the sample due to random fluctuation (i.e., near violations). Only domain knowledge allows us to differentiate them in practice, but their implications vary. Structural violations arise from an ill-defined target population, potentially due to a missing exclusion criterion \parencite{Platt_2012}, and no statistical procedure can fix them. Possible solutions are redefining the population of interest \parencite{Zhu_2021} or the potential outcomes involved \parencite{Nguyen_2022}. In contrast, random violations are more of an estimation issue that may be circumvented by techniques such as extrapolation \parencite{Leger_2022, Petersen_2012}. It is therefore crucial to qualitatively interpret the source of violations in order to assess their underlying cause.

Even if no studies investigated lack of positivity in mediation analysis, several studies demonstrated that several estimators of the total effect are biased when positivity is lacking \parencite{Leger_2022, Diop_2022, Moore_2012, Lendle_2013}.
When the goal is to estimate the total causal effect of an exposure $A$ on an outcome $Y$ (i.e., irrespective of any potential mediator), the positivity assumption means that each individual in the target population may theoretically receive each possible value of the studied exposure. However, in mediation analysis, the positivity assumption is more complex since it requires that the probabilities of each exposure and each mediator value of interest are non-zero \parencite{Imai_2010}. Furthermore, the specific form of the positivity assumption depends on the specific causal mediation estimand being targeted. \textcite{Nguyen_2022} formalized these positivity assumptions according to the different potential outcomes involved in the estimand of interest. Despite these critical developments, a gap remains: How can these positivity assumptions be verified in practice? 

Unfortunately, positivity has seldom been assessed in mediation settings. \textcite{Coffman_Zhong_2012} suggested estimating propensity scores for both the exposure and the mediator since extreme estimates indicate a lack of positivity. However, \textcite{Zhu_2021} illustrated that the propensity score can vary substantially depending on the model used, such that any evaluation of positivity may also substantially vary depending on analytical choices. Furthermore, extreme propensity scores may not identify the subgroups causing positivity violations; they only highlight specific individuals for whom the assumption appears to be violated based on the ensemble of their covariate values. 

For the total effect, \textcite{Danelian_2023} developed a method called Positivity Regression Trees (PoRT) to identify subgroups with practical positivity violations without requiring assumptions about the data-generating process or the specification of parametric models. In this study, we extend the PoRT approach from \textcite{Danelian_2023} to verify the practical positivity assumptions needed for various estimands of causal mediation analyses. We refer to this development as \textit{dePoRT} for \textit{decomposition PoRT}. 

The following section introduces our motivating example investigating the effect of childhood exposure to violence on dating violence in university students with various mediators. In the next two sections, we briefly present PoRT for the average causal effect (i.e., the total effect), after demonstrating why positivity is crucial for causal estimation. In Sections 4 to 6, we introduce dePoRT for the controlled direct, natural, and interventional effects, respectively. Finally, we discuss the implications of non-positivity in practice and outline strategies for addressing them.

\section{Motivating example}

Our motivating example is informed by a corpus of previous studies investigating the effect of childhood exposure to violence\footnote{Violence was defined broadly to include physical, psychological and sexual maltreatment.} ($A$) on dating violence perpetration ($Y$) mediated by religiosity involvement \parencite[$M$;][]{religiosity, Bierman_2005}. Post-traumatic stress symptoms (PTSS) are another mediator ($L$) of this exposure-outcome relationship in several populations \parencite[\textit{e.g.},][]{ptss, Jouriles_2012}, but are also posited as a mediator between childhood exposure to violence and religiosity \parencite{Leo_2021}. Figure \ref{dag} illustrates the posited causal structure.

\begin{figure}[htb]
\centering
\begin{tikzpicture}
    \node[circle, draw](C) at (0,1.5){$C$};
    \node[circle, draw](A) at (1.5,1.5){$A$};
    \node[circle, draw](M) at (3,1.5){$M$};
    \node[circle, draw](Y) at (4.5,1.5){$Y$};
    \node[circle, draw](L) at (2,0){$L$};
    
    \draw[->, thick, >=stealth]
    (C) edge (A)
    (A) edge (M)
    (M) edge (Y)
    (L) edge (M)
    (L) edge (Y)
    (C) edge [bend left=30] (M)
    (C) edge [bend left=30] (Y)
    (A) edge (L)
    (C) edge (L);

\end{tikzpicture}
\caption{Posited causal structure of the motivating example. $A$ is the childhood exposure to violence, $Y$ is the dating violence perpetration, $M$ is the religiosity involvement, $L$ is the post-traumatic stress symptoms, and $C$ is the set of confounders. \label{dag}}
\end{figure}

Our sample consists of university students from the United States of America, Canada, and Europe who have been in a relationship for at least one year. Among the 6,224 students included between 2001 and 2006, 1,175 (18.9\%) reported having been exposed to violence during childhood. This sample comes from the International Dating Violence Study \parencite{Straus_2011}. The study questionnaire includes two scales, the Conflict Tactics Scales \parencite{Straus_1996} measuring dating violence, and the Personal and Relationships Profile measuring several related risk factors \parencite{straus2010manual}. Confounders include gender, 
age, each parent's education level, family income, and parental marital status during childhood. We assume that all confounders are (accurately) recorded, and that there is no interference or selection bias.\footnote{See \textcite{Chatton_Rohrer_2024} for a discussion targeted towards psychologists about these issues.} Childhood exposure to violence is binary, while the scale for religiosity goes from 1 = none to 4 = high. Post-traumatic stress symptoms are measured as an average of several items bounded between 1 = none and 4 = high. Sample characteristics are presented in Appendix \ref{app:desc}.

\section{A journey in Positivity \label{sec:pos}}

To illustrate positivity, we will temporarily set aside the mediator and focus in this section on the total causal effect. As previously introduced, $A$ and $Y$ denote a binary exposure and an outcome, respectively. We use $Y^a$ to represent the potential outcome in a world where all individuals experienced the exposure $A=a$. We denote by $C$ the adjustment set used to achieve the exchangeability assumption, which will be defined shortly. This set should include confounders, but may include other variables. For example, pure causes of the outcome could be included to reduce the variance of various total and natural effects estimators \parencite{Brookhart_2006, Diop_2021, Chatton_2020}. The total effect is the contrast between the expectations of the two potential outcomes: $TE=\mathrm{E}(Y^1)-\mathrm{E}(Y^0)$. To identify this effect, we need three causal assumptions: 
\begin{seriate}
 \item consistency: $Y=Y^a$ if $A=a$,
 \item exchangeability: $Y^a \perp \!\!\! \perp A|C$, and
 \item positivity: $P(A=a|C=c) > \beta \geq 0, \forall c$ where $P(C=c)>0$
\end{seriate}
\parencite[p. 28]{Hernan_Robins_2020}. These assumptions allow mapping of the mean of potential outcomes $Y^a$ to a function of the observed outcomes and other data as follows:
\begin{align} 
\mathrm{E}(Y^a) &=  \sum_c\mathrm{E}(Y^a|C=c)\mathrm{P}(C=c) \\ 
 &=  \sum_c\mathrm{E}(Y^a|A=a, C=c)\mathrm{P}(C=c)\\
 &=  \sum_c\mathrm{E}(Y|A=a, C=c)\mathrm{P}(C=c). \label{gformula}
\end{align}
The first equality comes from the law of total expectation, the second from exchangeability and positivity, and the last from consistency. To understand why positivity is needed for the second equality, imagine a specific subgroup with $C=c^*$ in which no individual is experiencing the exposure $A = a$. In this case, $\mathrm{E}(Y^a|A=a, C=c^*)$ is undefined, and the second equality does not hold.\footnote{One can alternatively show identifiability using an inverse-probability-weighting expression. It puts the probability of the exposure in the denominator \parencite[p. 25]{Hernan_Robins_2020}, yielding a division by zero in case of lack of positivity.} Note, however, that we only need positivity for values of $c$ such that $\mathrm{P}(C=c)>0$. Indeed, when $\mathrm{P}(C=c)=0$ (with some abuse of notation or assuming discrete $C$), $\mathrm{E}(Y^a|A=a, C=c)\mathrm{P}(C=c) = 0$ regardless of whether $\mathrm{E}(Y^a|A=a, C=c)$ is well-defined or not. The range of values of $C$ for which the positivity assumption is required is called the ``support for positivity''. The positivity assumptions required for mediation analyses can be deduced similarly \parencite{Nguyen_2022} and are presented in Appendix \ref{proof}.

\section{Reach the PoRT}

\begin{figure}
    \centering
    \includegraphics[width=\linewidth]{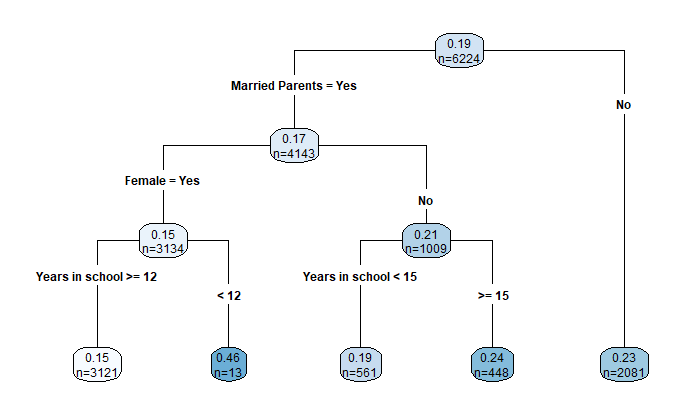}
    \caption{Example of a tree regression of childhood exposure to violence on three baseline confounders. The top node represents the whole sample, while the others each correspond to a subgroup defined by the previous splits. Each node's top number is the node-specific exposure prevalence.}
    \label{fig:rt}
\end{figure}

The PoRT algorithm is based on a succession of regression trees \parencite{Breiman_1984},\footnote{While the dependent variable in the tree is categorical, PoRT uses regression trees instead of classification trees. Both are valid; their difference lies in the loss function used to split the data: the Gini index and the sum of squared errors of the child nodes, respectively. Because regression trees operate on predicted probabilities, they may retain a split that affects predicted probabilities even if the split does not affect the predicted class, unlike classification trees. See \textcite{Strobl_2009} for an introduction to regression trees in psychology.
} where the exposure is regressed on one or more confounders. Regression trees split the data into subgroups (called nodes) according to successive binary partitions learned from the data (Figure \ref{fig:rt}). Importantly, regression trees can learn nonlinear and nonmonotone association rules from the data, without the need to specify them in advance, unlike parametric regressions. These splits lead to potentially interpretable subgroups (i.e., nodes representing subgroups with homogeneous exposure levels), which explains the popularity of this method for prediction purposes. While we are not interested in prediction, the interpretability of the resulting subgroups is the heart of PoRT. For instance, the bottom-left node in Figure \ref{fig:rt} represents the subgroup of individuals having married parents, identifying themselves as female, and being at school since at least 12 years.

PoRT works as follows. It takes four inputs: \begin{seriate}
    \item confounder set $C$,
    \item values of two flagging hyperparameters  $\alpha$ and $\beta$, and
    \item value of stopping hyperparameter $\gamma$.
\end{seriate} 
First, for each confounder, a tree is built using the exposure as the dependent variable and the confounder as the sole independent variable. Second, each node is ``read'', with respect to the chosen values of $\alpha$ and $\beta$, to identify positivity violations (described below). If a violation is found, the confounder is removed from $C$ for future iterations of PoRT. The same process is then repeated, but two confounders are used together as independent variables. This process is iterated by progressively increasing the number of independent variables considered simultaneously until the value of the stopping hyperparameter $\gamma$ is exceeded. For instance, if $\gamma=3$, one tree per confounder is fit, then one tree per pair of remaining confounders, and finally, one tree per trio of remaining confounders.

But how to ``read'' these trees? The goal is to identify subgroups (i.e., nodes) violating the empirical bound of the positivity assumption; these correspond to nodes where the probability of exposure is lower than a bound $\beta$. This bound represents a threshold where the exposure probability becomes too small for the exposure causal effect in a subgroup to be supported by the data \parencite{DAmour_2021}. \textcite{Gruber_2022} proposed a mean square error-optimal bound of $\beta=5/(\! \sqrt{n} \cdot \ln{(n)})$ to truncate propensity score weights to circumvent practical non-positivity. The value of $\beta$ must be specified \textit{a priori} as a hyperparameter of PoRT. The second hyperparameter $\alpha$ controls the size of identified subgroups by excluding the smallest ones (for instance, those consisting of less than 1\% of the total number of individuals from the study sample). Positivity violations in very small subgroups may not impact the estimation when the model extrapolates over the subgroup \parencite{Petersen_2012, Leger_2022}. 

In our motivating example, positivity for the exposure must be checked for any potential outcome involved in a causal effect \parencite{Nguyen_2022}. Therefore, we must check positivity related to the identifiability of $\mathrm{E}(Y^1)$ and $\mathrm{E}(Y^0)$. First, we fit a regression tree conditional on each confounder. With $\alpha=0.01$, $\gamma=2$, and Gruber's bound for $\beta$, we found no subgroups that have both (i) more than 62 individuals (i.e., more than $\alpha = $1\% of the total sample size) and (ii) a proportion exposed to childhood violence below Gruber's bound of $\beta = $0.7\%. Then, a tree with each pair of confounders was built (since no violations were identified in the first step, no confounders were excluded in this second step) and read identically. Again, no violations were identified. No more complex trees were fit since the hyperparameter $\gamma=2$ was reached. In summary, no violations of the positivity assumption for the total effect estimation were found. Since positivity was found to hold, we proceeded with the estimation of the total effect. Here, we used the inverse probability weighting estimator \parencite{Robins_2000}, which is well known in the psychological community \parencite{Harder_2010, Lanza_2013}. We thus obtained an odds ratio (OR) of 1.81 with a 95\% confidence interval (95\% CI) from 1.51 to 2.16.

We now turn to dePoRT for various mediational effects. 

\section{Simplest case: Controlled direct effect}

\subsection{Positivity}
Controlled direct effects (CDEs) are causal effects of the exposure in a world where the mediator is set to a specific level $m$ for everyone. Formally, $CDE(m) = \mathrm{E}(Y^{1,m}) - \mathrm{E}(Y^{0,m})$, where $Y^{a,m}$ stands for the outcome in a counterfactual world where all individuals are exposed to $A=a$ and $M=m$. Positivity for $\mathrm{E}(Y^{a,m})$ implies both $\textrm{P}(A=a|C=c)>\beta\geq0$ and $\textrm{P}(M=m|A=a,C=c)>\beta\geq0, \forall c$ such that $P(C=c)>0$ \parencite{Nguyen_2022}.

\subsection{dePoRT}
These two positivity assumptions for the CDE are similar in form to the positivity assumption for the total causal effect. As such, dePoRT uses the exact same PoRT algorithm as for the total effect thrice: once regressing the exposure $A$ as the dependent variable and the covariates $C$ as the independent variables, and then regressing the mediator $M=m$ as the dependent variable and the covariates as the independent variables, while subsetting on $A=a$ for each of $a=0$ and 1. For the latter application of PoRT, note that the mediator needs to be dichotomized in order to model $M = m$ vs $M \neq m$ for the specific $m$ of interest.

\subsection{Application}
Let's illustrate positivity evaluation for CDE through our motivating example. Imagine we study the effect of childhood exposure to violence ($A$) on dating violence perpetration ($Y$) while setting their religious involvement ($M$) to a high level ($m = 4$). Note that we do not consider PTSS as a mediator until the section related to interventional effects for pedagogical purposes. Positivity for the exposure was already checked in the previous section. We kept the same hyperparameter values (\textit{i.e.}, $\alpha=0.01$ and $\gamma=2$) for exposure and mediational positivities. Since we used Gruber's bound for $\beta$, its value differs across the two subsets due to their different sizes ($\beta=0.008$ for individuals with $A=0$ and $\beta=0.021$ for individuals with $A=1$). Applying PoRT to estimate the probability of high religious involvement (using a dichotomous version of the original ordinal variable $M$; i.e., the highest value vs the others) according to the childhood exposure to violence and the covariates identifies 14 subgroups of individuals for whom the mediator positivity assumption is violated. One of these subgroups concerns students who were not exposed to violence during childhood, while thirteen subgroups concern students who were (Table \ref{tab:cde}). As an example, the result for the first problematic subgroup among the exposed in Table \ref{tab:cde} should be interpreted as follows: The 57 individuals with a childhood exposure to violence ($A=1$) with both (i) a father education z-score between -2 and -0.5 and (ii) a mother education z-score between -1.5 and -1 present a high religious involvement ($M=4$) probability of 2\%, which is lower than the posited positivity threshold (here 2.1\%). Consequently, the positivity assumption is deemed violated for those individuals. Note that we are applying the $\alpha$ and $\beta$ thresholds directly to the sample size of the exposure group. The final step would be to determine whether the identified violations are structural or random, based on subject-matter expertise. Here, it seems unlikely that this violation is structural, as a high religious involvement is theoretically possible regardless of parents' education. 

Because there are no structural violations, we can begin the estimation. The presence of a few random violations may lead us to prefer a causal estimator that can better extrapolate over regions where there are positivity violations \parencite{Petersen_2012}. Hence, we used the iterated g-formula \parencite{Robins_1986, Loh_2024} that is implemented in the CMAverse \texttt{R} package \parencite{Shi_2021} for mediation analyses instead of an inverse probability weighting estimator. We obtained an OR of 1.87 (95\% CI from 1.61 to 2.22).

\begin{table}[h!]
\begin{center}
\fontsize{10}{15}\selectfont
\begin{threeparttable}
\caption{Mediator positivity violations for the controlled direct effect application\label{tab:cde}
}
\begin{tabular}{lcc}
  \hline
 Subgroup & Prob. & $n^* (\%)$ \\ 
  \hline
  \multicolumn{3}{l}{$P(M=4 | A=1, C)$, $\beta=0.021$}\\
  FEZ $\in \rinterval{-2}{-0.5}$ \& MEZ $\in \rinterval{-1.5}{-1}$ & 0.02 &  57 (4.9) \\ 
  FEZ $\in \rinterval{-1}{-0.5}$ \& MEZ $\in \rinterval{0.5}{1.5}$ & 0.00 &  22 (1.9) \\ 
  Age $\in \rinterval{25}{30}$ \& FEZ $\in \rinterval{-2}{-1}$ & 0.00 &  17 (1.4) \\ 
  FEZ $\in \rinterval{-2}{-1.5}$ \& Age $\in \rinterval{23}{25}$ & 0.00 &  13 (1.1) \\ 
  Age $\in \rinterval{25}{30}$ \& FEZ $\in \rinterval{0.5}{1}$ & 0.00 &  12 (1.0) \\ 
  MEZ $\in \rinterval{-1.5}{-1}$ \& FIZ $< -1$ & 0.00 &  24 (2.0) \\ 
  Age $<30$ \& MEZ $\in \rinterval{-2}{-1.5}$ & 0.00 &  34 (2.9) \\ 
  MEZ $\in \rinterval{-1.5}{-1}$ \& Age $\in \rinterval{21}{27}$ & 0.00 &  26 (2.2) \\ 
  Age $\in \rinterval{27}{30}$ \& MEZ $\in \rinterval{-1}{-0.5}$ & 0.00 &  16 (1.4) \\ 
  Age $\in \rinterval{30}{45}$ \& MEZ $\in \rinterval{-1.5}{-1}$ & 0.00 &  13 (1.1) \\ 
  FIZ $\in \rinterval{-1}{0}$ \& Age $\in \rinterval{19}{20}$ & 0.02 &  51 (4.3) \\ 
  Age $\in \rinterval{25}{30}$ \& FIZ $<-1$ & 0.00 &  29 (2.5) \\ 
  &&\\
  \multicolumn{3}{l}{$P(M=4 | A=0, C)$, $\beta=0.008$}\\
  MEZ $< -2.5$ & 0.00& 56 (1.1)\\
   \hline
\end{tabular}
\begin{tablenotes}
\item Abbreviations: $\beta$, Gruber bound; FEZ, Father education z-score; FIZ, Family income z-score; MEZ, Mother education z-score; \textit{n*}, subgroup size; Prob., Estimated probability of interest. 
\end{tablenotes}
\end{threeparttable}
\end{center}
\end{table}

\section{Decomposing causal pathways: Natural effects}

Consider now natural effects that allow us to decompose the total effect of the exposure on the outcome into direct and indirect effects through a mediator. The potential outcomes are defined in a hypothetical world where individuals experience the exposure $A=a$, but the mediator takes the value it would have under the opposite exposure $A=a^*$, formally $Y^{a,M^{a^*}}$. As we will see, this requires different positivity assumptions.

\subsection{Positivity}

The focus is on the decomposition of the total effect on either pure natural direct effect and total natural indirect effect (formally $PNDE=\mathrm{E}(Y^{1,M^{0}}) - \mathrm{E}(Y^{0})$ and $TNIE=\mathrm{E}(Y^{1}) - \mathrm{E}(Y^{1,M^{0}})$) or pure natural indirect effect and total natural direct effect (formally $PNIE=\mathrm{E}(Y^{0,M^{1}}) - \mathrm{E}(Y^{0})$ and $TNDE=\mathrm{E}(Y^{1}) - \mathrm{E}(Y^{0,M^{1}})$) \parencite{MacKinnon_2020}. Here, the challenging potential outcome's expectation is $\mathrm{E}(Y^{a,M^{a^*}})$, where $a \ne a^*$.\footnote{Natural effects could also be defined by replacing $\mathrm{E}(Y^{a})$ with $\mathrm{E}(Y^{a,M^a})$. However, under the composition assumption $Y^{a,M^{a}}=Y^a$ both expressions are equivalent \parencite[p. 229]{Pearl_2000}.} 

\textcite{Pearl_2001} derived the so-called mediation formula for natural effects as \begin{multline}\label{medform} \mathrm{E}(Y^{a,M^{a^*}})=  \sum_m\sum_c\mathrm{E}(Y|A=a, M=m, C=c) \times\\ \mathrm{P}(M=m|A=a^*,C=c)\mathrm{P}(C=c) \end{multline} (see Appendix \ref{proof} for the proof). 

The positivity assumption encompasses both the exposure and the mediator. Contrary to the CDE, natural effects involve a nested potential outcome's expectation $\mathrm{E}(Y^{a,M^{a^*}})$ that implies exposure positivity as both $\mathrm{P}(A=a|C=c)>\beta \geq 0$ and $\mathrm{P}(A=a^*|C=c)>\beta \geq 0$. One may illustrate this through Equation \eqref{medform}. The quantity $\mathrm{E}(Y|A=a, M=m, C=c)$ is undefined when $\mathrm{P}(A=a|C=c)>\beta \geq 0$, as discussed in the Section ``A journey in Positivity'', while $\mathrm{P}(M=m|A=a^*,C=c)$ is undefined when $\mathrm{P}(A=a^*|C=c)>\beta \geq 0$ by applying the same reasoning.
Furthermore, positivity of the mediator must hold, within each observed
stratum $c$, for each value $m$ observed in the opposite counterfactual world (i.e., where $A=a^*$) \parencite{Nguyen_2022}:  $\mathrm{P}(M=m|A=a,C=c)>\beta\geq0, \forall m \ s.t. \ \mathrm{P}(M=m|A=a^*,C=c)>0$.
Mediator positivity is lacking when, focusing on the particular subgroup $C=c$, the mediator values' range is more restricted among those with $A=a$ than among those with $A=a^*$. When this happens, the mediator positivity is respected for the opposite cross-world outcome's expectation $\mathrm{E}(Y^{a^*,M^{a}})$ for this subgroup $C = c$, since the support of the mediator's distribution among $A=a^*$ fully covers that among $A=a$. Therefore, it's possible that positivity violations could lead to one decomposition (i.e., PNDE and TNIE vs TNDE and PNIE) being identified when another is not. Details, including a toy simulation study, are given in Appendix \ref{app:decomp}.

\subsection{dePoRT}

dePoRT is more challenging to apply to nested potential outcome's expectations, such as $\mathrm{E}(Y^{a,M^{a^*}})$.
Indeed, one must now check, for all possible covariate values $c$, whether the mediator's distribution in the exposed covers those of the unexposed individuals.
We suggest using a ``reverse engineering'' trick as follows. First, apply PoRT (i) with the mediator as the dependent variable, and (ii) only among individuals with $A=a$ to find subgroups $c'$ where the mediator-related part of the positivity assumption is potentially violated. Second, for each problematic subgroup $c'$ identified this way, compute the mediator prevalence among individuals with $A=a^*$ and $C = c'$. If the resulting prevalence is lower than $\beta$, the potential positivity violation occurs in a subgroup that can be deemed outside of the support of interest and is therefore not an actual positivity violation; it can be ignored. If the resulting prevalence is greater than $\beta$, an actual positivity violation is present in the subgroup $c'$. The exposure positivity is checked as usual. The whole process is implemented in the \texttt{R} package \texttt{port} \parencite{port_package} through the function \texttt{deport}, and illustrated in the notebook available at github.com/ArthurChatton/dePoRT-notebook.

\subsection{Application}

\begin{table*}[h!]
\begin{center}
\begin{minipage}{2\textwidth}
\begin{threeparttable}
\caption{Mediator positivity violations for natural effects \label{tab:ne}
}
\begin{tabular}{lccccc}
  \hline
 Subgroup $C=c$ & $m$ & Prob. & $n^* (\%)$  & Prev. $a^*$ & Violation for\\ 
  \hline
  \multicolumn{6}{l}{$P(M=m | A=a=0, C)$, $\beta=0.008$}\\
 Age = 23 \& MEZ $\in \rinterval{1.5}{2.5}$  & 3  & 0.00 & 51 (1.0) & 0.00 & None    \\
 \multicolumn{6}{l}{}\\
  \multicolumn{6}{l}{$P(M=m | A=a=1, C)$, $\beta=0.021$}\\
  Age $\geq$ 45 \& FEZ $<-1$   & 1 & 0.00 &  16 (1.4) & 0.18 & $\mathrm{E}(Y^{1,M^0})$ \\
   FEZ $<-2$ \& FIZ $\geq -1$ & 1 & 0.00 & 25 (2.1) & 0.20 & $\mathrm{E}(Y^{1,M^0})$\\
  FEZ $\in \rinterval{-0.5}{0.5}$ \& MEZ $\geq 1.5$ & 3 & 0.00 & 13 (1.1) & 0.05 & $\mathrm{E}(Y^{1,M^0})$\\
   Age $\geq 36$ \& FEZ $\geq 0.5$  & 3 & 0.00 & 15 (1.3) & 0.06 & $\mathrm{E}(Y^{1,M^0})$\\
   Age $\in \rinterval{30}{36}$ \& MEZ $\in \rinterval{-1}{-0.5}$ & 3 & 0.00 & 19 (1.6) & 0.10 & $\mathrm{E}(Y^{1,M^0})$\\
   Age $\geq 45$ \& FIZ $\in \rinterval{0}{1}$  & 3 & 0.00 & 20 (1.7) & 0.07 & $\mathrm{E}(Y^{1,M^0})$\\
   FEZ $\in \rinterval{-2}{-1.5}$ \& MEZ $\geq -0.5$ & 4 & 0.00 & 14 (1.2) & 0.14 & $\mathrm{E}(Y^{1,M^0})$\\
   \hline
\end{tabular}
\begin{tablenotes}
\vspace{0.1cm}
\item Abbreviations: $\beta$, Gruber bound; FEZ, Father education z-score; FIZ, Family income z-score; MEZ, Mother education z-score; \textit{n*}, subgroup size; Prob., Estimated probability of interest; Prev., Estimated prevalence of interest within the alternative exposure group (i.e., $A=a^*$). 
\end{tablenotes}
\end{threeparttable}
\end{minipage}
\end{center}
\end{table*}

Let's return to our motivating example, but now consider the entire range of the mediator's values (religious involvement on a 1-4 psychometric scale). We still use the previous set of hyperparameters.  

For checking mediator positivity, we begin by running PoRT using the mediator religiosity as the dependent variable in the decision trees (see Section ``Reach the PoRT'' for verification of exposure positivity). Since religiosity is no longer binary, yet still categorical, classification trees are used instead of regression trees, with no impact on interpretation. For $\mathrm{E}(Y^{0,M^1})$, the trees are built on the subset of individuals with $A=0$. Only one violation was identified at this step and presented at the top of Table \ref{tab:ne}. Now, we compute the prevalence of the problematic religiosity category ($M=3$) in the same subgroup (i.e., 23-year-old students with a mother's education z-score between 1.5 and 2.5) but among individuals with $A=1$; this prevalence is equal to 0. Since the mediator value was not observed in this subgroup among $A=1$, it is not needed for the same subgroup definition among $A=0$. Thus, the subgroup does not violate the mediator positivity for $\mathrm{E}(Y^{0,M^1})$. 

Second, we shift the exposure groups. Mediator positivity for $\mathrm{E}(Y^{1,M^0})$ is checked using PoRT on the individuals with $A=1$ first. Ten violations were identified at this step (bottom of Table \ref{tab:ne}). The prevalence of these subgroups is now checked in the other exposure group (i.e., $A=0$), and all subgroups showed a prevalence higher than $\beta$, confirming mediator positivity violations.

In this application, we may prefer to use the decomposition involving the nested potential outcome $\mathrm{E}(Y^{0,M^1})$ (i.e., $TNDE$ and $PNIE$) since it presents a lower risk of bias related to mediator positivity violations than its \textit{alter ego} (i.e., $TNIE$ and $PNDE$). 

For the estimation, we used a weighting-based natural model \parencite{Lange_2012}, implemented in the medflex \texttt{R} package \parencite{medflex}. On the OR scale, the TNDE was estimated to be 1.79 (95\% CI from 1.48 to 2.18) while the PNIE estimate was 1.01 (95\% CI from 0.99 to 1.03). These results indicate that exposure to violence during childhood impacts conjugal violence, but not through religiosity. However, natural effects are not the appropriate effects to look for in this application since PTSS is a post-exposure confounder, as detailed in the next section.

\section{Post-exposure confounding: Interventional effects}

Until now, we have set aside PTSS, a mediator-outcome confounder influenced by the exposure. However, in the presence of such an intermediate confounder $L$ (as illustrated in Figure \ref{dag}), natural effects are not identifiable \parencite{Avin_2005}. Interventional mediation effects represent a solution to this problem \parencite{VanderWeele_Vansteelandt_2014}.

Interventional effects are often used to try to mimic how natural effects decompose the total effect into direct and indirect effects.\footnote{Interventional effects are sometimes referred to as the randomized-intervention analogue of natural effects \parencite{VanderWeele_Vansteelandt_2014} or as stochastic effects \parencite{Diaz_Hejazi_2020}. However, note that they do not truly decompose the total effect, but rather an undefined overall effect \parencite{VanderWeele_Vansteelandt_Robins_2014}. Sometimes natural and interventional effects coincide, notably when there is no interaction between $L$ and $M$ in the true data-generating process \parencite{Gervais_2025}.} Interventional effects consider counterfactual outcomes under interventions on both the exposure and the mediator, similar to the CDE. However, the intervention on the mediator is stochastic (i.e., it has a random component). The goal is for the intervention on the mediator to emulate the population distribution of the mediator under fixed exposure $A=a^*$ and conditional on covariates $C$.\footnote{\textcite{Nguyen_2022} also discuss a draw from a distribution of $M^{a^*}$ conditional on ($C$, $L$) representing an intervention definition that might be plausibly linked to a real-life intervention.} The distribution of the mediator under this stochastic intervention is denoted hereafter by $\mathcal{M}^{a|C}$ and $\mathcal{M}^{a^*|C}$, following \textcite{Nguyen_2022}. The interventional direct effect is often expressed as $PIDE$ = $\textrm{E}(Y^{1, \mathcal{M}^{0|C}}) - \textrm{E}(Y^{0, \mathcal{M}^{0|C}})$, and the interventional indirect effect as $TIIE$ = $\textrm{E}(Y^{1, \mathcal{M}^{1|C}}) - \textrm{E}(Y^{1, \mathcal{M}^{0|C}})$, similar to the decomposition into a $PNDE$ and a $TNIE$. It is, however, also possible to express the interventional direct effect as $TIDE$ = $\textrm{E}(Y^{1, \mathcal{M}^{1|C}}) - \textrm{E}(Y^{0, \mathcal{M}^{1|C}})$ and the interventional indirect effect as $PIIE$ = $\textrm{E}(Y^{0, \mathcal{M}^{1|C}}) - \textrm{E}(Y^{0, \mathcal{M}^{0|C}})$, similar to the decomposition into a $TNDE$ and a $PNIE$.\footnote{Note that the composition assumption does not hold for interventional effects since $Y^{a,\mathcal{M}^{a|C}}$ and $Y^{a}$ are not equivalent \parencite{Nguyen_2021}.}

\subsection{Positivity}

Under appropriate identifiability assumptions, the potential outcome means are identified as follows \parencite{Nguyen_2022}: 
\begin{multline}\label{gf_int}
\mathrm{E}(Y^{a,\mathcal{M}^{a|C}}) = \sum_{c,l,m} \mathrm{E}(Y|a,m, l,c) \times\\\mathrm{P}(M=m|a,c)\mathrm{P}(L=l|a,c)\mathrm{P}(C=c)
\end{multline}
\begin{multline}\label{gf_int_nested}
\mathrm{E}(Y^{a,\mathcal{M}^{a^*|C}}) = \sum_{c,l,m} \mathrm{E}(Y|a,m, l,c) \times\\\mathrm{P}(M=m|a^*,c)\mathrm{P}(L=l|a,c)\mathrm{P}(C=c)
\end{multline}
where we use lowercase letters only on the right side of the conditional probabilities to alleviate the notational burden.

Following \eqref{gf_int} and contrary to natural effects, mediator positivity is now required for the potential outcomes' expectations $\mathrm{E}(Y^{a,\mathcal{M}^{a|C}}) $. Formally, mediator positivity is defined as $\mathrm{P}(M=m|A=a,L=l,C=c)>\beta \geq0, \forall m,l$ such that $\mathrm{P}(M=m|A=a,C=c)>0$ and $\mathrm{P}(L=l|A=a,C=c)>0$. Positivity for nested potential outcomes' expectations $\mathrm{E}(Y^{a,\mathcal{M}^{a^*|C}}) $ is similar to that of natural effects. The only difference is that mediator positivity is now $\mathrm{P}(M=m|A=a,L=l,C=c)>\beta \geq 0$ instead of $\mathrm{P}(M=m|A=a,C=c)>\beta \geq 0$. However, the constraint on the support is the same, i.e. mediator positivity is needed only for values $m$ such that $\mathrm{P}(M=m|A=a^*,C=c)>0$.

\subsection{dePoRT}

The diagnostic procedure is similar to the one we used for natural effects, with some notable differences arising from the definition of the support. 

For $\mathrm{E}(Y^{a,\mathcal{M}^{a|C}})$, the first step consists of running PoRT for exposure positivity as usual. The second step is to run PoRT for mediator positivity on each exposure group $A=a$ separately, conditional on $(C, L)$. Potential violations, for a given $m$, identified at this step need to be confirmed in a third step by verifying whether they occur on the support, that is, in regions of the covariate space where $\mathrm{P}(M=m|A=a,C=c)>0$ and $\mathrm{P}(L=l|A=a,C=c)>0$. For example, suppose we identify the following potential positivity violation: $\mathrm{P}(M=m|A=1, L=l', C=c')\leq\beta$, where $c'$ and $l'$ stand for specific values of $C$ and $L$, respectively. The third step is now to compute the prevalence of $m$ among individuals with $A=1$ and $C=c'$, regardless of their $L$ value. There is no violation if the prevalence computed at this third step is null. The special case where the subgroup involves $L$ but not $C$ (e.g., $\mathrm{P}(M=m|A=1,L=l')\leq\beta$) does not represent a positivity violation either since the support here would be the whole exposure group.

The process is almost identical for $\mathrm{E}(Y^{a,\mathcal{M}^{a^*|C}})$, except that one computes the prevalence of the mediator in the problematic subgroup on the other exposure group at step 3. Formally, for a violation observed in exposure group $A=a$, one computes $\mathrm{P}(M=m|A=a^*,C=c')$ rather than $\mathrm{P}(M=m|A=a,C=c')$. As for natural effects, the varying amount of mediator positivity violations across the nested potential outcomes may influence the choice of decomposition of interest before conducting any statistical analysis.

\subsection{Application}

Our motivating example still includes a binary exposure (childhood exposure to violence), a four-class mediator (religious involvement), and the following hyperparameters: $\alpha=0.01$, $\beta=5/(\! \sqrt{n} \cdot \ln{(n)})$, and $\gamma=2$. But, this time, it also includes an intermediate confounder (PTSS) on a continuous scale. As previously, we focus on mediator positivity since exposure positivity was already explained in Section ``Reach the PoRT''.

\begin{table*}[ht!]
\begin{center}
\begin{minipage}{2\textwidth}
\begin{threeparttable}
\caption{Mediator positivity violations for interventional effects \label{tab:ie}
}
\begin{tabular}{lcccccc}
  \hline
 Subgroup $\{C=c, L=l\}$ & $m$ & Prob. & $n^* (\%)$  & Prev. $a$ & Prev. $a^*$ & Violation for\\ 
  \hline
  \multicolumn{6}{l}{$P(M=m | A=a=0, C, L)$, $\beta=0.008$}\\
 Age = 23 \& MEZ $\in \rinterval{1.5}{2.5}$  & 3  & 0.00 & 51 (1.0) & 0.00 & 0.00 & None    \\
 PTSS $\in \rinterval{1.5}{1.75}$ \& MEZ $\in \rinterval{-1.5}{-1}$  & 3  & 0.00 & 51 (1.0) & 0.05 & 0.13 & $\mathrm{E}(Y^{0,\mathcal{M}^{0|C}}), \mathrm{E}(Y^{0,\mathcal{M}^{1|C}})$    \\
 \multicolumn{6}{l}{}\\ 
  \multicolumn{6}{l}{$P(M=m | A=a=1, C, L)$, $\beta=0.021$}\\
  PTSS $\in \rinterval{3.5}{3.625}$   & 4 & 0.00 &  14 (1.2) & - & - & None \\
   FEZ $\in \rinterval{-2}{-1.5}$ \& MEZ $\geq -0.5$ & 4 & 0.00 & 14 (1.2) & 0.00 & 0.14 & $\mathrm{E}(Y^{1,\mathcal{M}^{0|C}})$\\
    FEZ $\in \rinterval{-0.5}{0.5}$ \& MEZ $\geq 1.5$ & 3 & 0.00 & 13 (1.1) & 0.00 &  0.05 & $\mathrm{E}(Y^{1,\mathcal{M}^{0|C}})$\\
   FEZ $< -2$ \& FIZ $\geq -1$ & 1 & 0.00 & 25 (2.1) & 0.00 & 0.20 & $\mathrm{E}(Y^{1,\mathcal{M}^{0|C}})$\\ 
    Age $\geq 36$ \& FEZ $\geq 0.5$  & 3 & 0.00 & 15 (1.3) & 0.00 & 0.06 & $\mathrm{E}(Y^{1,\mathcal{M}^{0|C}})$\\
     Age $\geq 45$ \& FEZ $<-1$  & 1 & 0.00 & 16 (1.4) & 0.00 & 0.18 & $\mathrm{E}(Y^{1,\mathcal{M}^{0|C}})$\\
    Age $\rinterval{30}{36}$ \& MEZ $\in \rinterval{-1}{-0.5}$  & 3 & 0.00 & 19 (1.6) & 0.00 & 0.10 & $\mathrm{E}(Y^{1,\mathcal{M}^{0|C}})$\\
  Age $\geq 45$ \& FIZ $\in \rinterval{0}{1}$ & 3 & 0.00 & 20 (1.7) & 0.00 & 0.07 & $\mathrm{E}(Y^{1,\mathcal{M}^{0|C}})$\\
   \hline
\end{tabular}
\begin{tablenotes}
\vspace{0.1cm}
\item Abbreviations: $\beta$, Gruber bound; FEZ, Father education z-score; FIZ, Family income z-score; MEZ, Mother education z-score; \textit{n*}, subgroup size; Prob., Estimated probability of interest; Prev., Estimated prevalence of interest in the comparator group (\textit{i.e.}, without $L$ and within $a$ for $\mathrm{E}(Y^{a,\mathcal{M}^{a|C}})$ or $a^*$ for $\mathrm{E}(Y^{a,\mathcal{M}^{a^*|C}})$); PTSS, Post-traumatic stress disorder. 
\end{tablenotes}
\end{threeparttable}
\end{minipage}
\end{center}
\end{table*}

First, we ran PoRT on the exposed individuals to (i) obtain a classification tree modeling the religious involvement conditional on each confounder (including PTSS), (ii) read them to save potential positivity violations given $\alpha$ and $\beta$, (iii) exclude the confounder(s) involved in these trees, (iv) build other classification trees using each possible pair of the remaining confounders, (v) read them to save the potential positivity violations. This iterative process stops when the number of confounders per tree exceeds $\gamma$ at the next iteration. Second, we apply the very same process to the unexposed individuals. 

Following this procedure, we obtained two lists of subgroups that yield potential positivity violations. Our goal is to reduce them according to the support that defines the four potential outcomes. For $\mathrm{E}(Y^{1,\mathcal{M}^{1|C}})$ and $\mathrm{E}(Y^{0,\mathcal{M}^{0|C}})$, we pick the subgroups involving any value or range of PTSS. We compute the prevalence of the mediator value that lacks positivity, but among the broader subgroup that does not condition on PTSS (for exposed or unexposed, respectively). For example, the subgroup of unexposed individuals with both a PTSS between 1.5 and 1.75 and a mother education z-score between -1.5 and -1 was identified at the first step (no $m=3$ in this subgroup, Table \ref{tab:ie}). The prevalence of $m=3$ among unexposed individuals who have only a mother's education z-score between 1.5 and -1 (regardless of their PTSS value) is 0.05. Since this last prevalence is greater than $\beta$, the subgroup represents a mediator positivity violation for $\mathrm{E}(Y^{0,\mathcal{M}^{0|C}})$. 

The process is similar for $\mathrm{E}(Y^{1,\mathcal{M}^{0|C}})$ and $\mathrm{E}(Y^{0,\mathcal{M}^{1|C}})$, except that one checks the prevalence in the broader subgroup among the other exposure group. Back to our problematic subgroup, we now compute the prevalence of $m=3$ among exposed individuals (rather than unexposed) that have only a mother's education z-score between 1.5 and -1 (still regardless of their PTSS value). This value being greater than $\beta$, the subgroup also represents a mediator positivity violation for $\mathrm{E}(Y^{0,\mathcal{M}^{1|C}})$. In contrast, the subgroup of 23-year-old individuals with a mother's education z-score (potential violation for $\mathrm{E}(Y^{0,\mathcal{M}^{1|C}})$, first line of Table \ref{tab:ie}) also presented a null prevalence among the exposed, negating the violation. To summarize, a subgroup $(C=c', L=l')$ identified by dePoRT may represent a mediator positivity violation (for the value $M=m$) for $\mathrm{E}(Y^{a,\mathcal{M}^{a|C}})$ if $\mathrm{P}(M=m|A=a,C=c')\geq\beta$ and for $\mathrm{E}(Y^{a,\mathcal{M}^{a^*|C}})$ if $\mathrm{P}(M=m|A=a^*,C=c')\geq\beta$.

Looking at the complete results in Table \ref{tab:ie}, we have one mediator positivity violation for $\mathrm{E}(Y^{0,\mathcal{M}^{0|C}})$ and $\mathrm{E}(Y^{0,\mathcal{M}^{1|C}})$, and ten others for $\mathrm{E}(Y^{1,\mathcal{M}^{0|C}})$. In this situation, one may prefer using the decomposition based on $\mathrm{E}(Y^{0,\mathcal{M}^{1|C}})$, to reduce the risk of bias related to non-positivity.

We use the g-formula estimator from the CMAverse \texttt{R} package \parencite{Shi_2021}. The TIDE, on an OR scale, is 1.83 (95\% CI from 1.50 to 2.23), while the PIIE is estimated at 0.99 (95\% CI from 0.98 to 1.01), confirming the previous conclusions about the absence of a mediating role for religiosity involvement.

\section{Discussion}

Causal mediation analyses aim to answer highly relevant scientific questions, but their assumptions are especially strong \parencite{Rohrer_2022, Keele_2015} and their application is challenging. Unfortunately, these assumptions are rarely assessed in mediation analysis \parencite{Stuart_2021}, likely due to a lack of tools to check them. In this study, we propose the dePoRT algorithm to check the positivity assumption in causal mediation analysis. Built as a wrapper around PoRT, dePoRT facilitates both the assessment and interpretation of the positivity assumption in causal mediation analysis. 

We illustrated the use of dePoRT with the potential outcomes involved in defining three common estimands: the CDE, natural effects, and interventional effects. The relevance of interventional effects is currently debated in the literature, with pro and con arguments \parencite[see e.g.,][]{MorenoBetancur_Carlin_2018, Miles_2023, Loh_Ren_2022}. Other estimands might be of interest when intermediate confounders are present. For instance, \textcite{Chen_Lin_2025} introduced separable path-specific effects while  \textcite{Gervais_tuto_2025} discussed the joint effect of $L$ and $M$ as well as natural path-specific effects. The positivity of the mediator will differ for each of them. We can apply the same reasoning to check them using PoRT: derive the mediation formula, define the mediator positivity and the support of $m$, and finally build a specific version of dePoRT using the \texttt{port} function \parencite{port_package}. Finally, positivity for the causal contrast of interest (e.g., natural effects) needs exposure and mediator positivities for all potential outcomes involved in this contrast. We chose to check the positivity involved in the identifiability of different expectations of the potential outcomes (e.g., $\mathrm{E}(Y^a)$ or $\mathrm{E}(Y^{a,M^{a^*}})$) for two reasons. First, it allows for studying the different decompositions. This is also useful when the interest lies in path-specific natural effects with several mediators. Second, it can highlight the non-identifiability of some causal contrasts and may lead to choosing a less realistic but still informative causal estimand. For instance, \textcite{Nguyen_2022} discussed a modified intervention defined by the difference between $\mathrm{E}(Y^{a,\mathcal{M}^{a^*|C}})$ and $\mathrm{E}(Y^{a^*})$ instead of the PIIE considered in the section related to interventional effects.

In this study, we mostly set aside the estimation process. Nonetheless, some weighting estimators, such as inverse odds ratio weighting \parencite{Tchetgen_2013}, use weights that have a denominator estimated by a model of the exposure conditional on the mediator. In such cases, dePoRT could be used to confirm whether the data are sufficient to support the estimation of the weights. We also only considered binary exposure and categorical mediators. Contrary to \textcite{Danelian_2023}, we employed classification trees instead of regression trees to handle non-binary categorical mediators (this approach could also be applied to the exposure) since the latter has not yet been implemented in the \texttt{R} package \texttt{rpart} \parencite{Therneau_Atkinson} for a multi-class dependent variable. Continuous exposures or mediators are more challenging to deal with. One can typically expect a lack of positivity in such cases because the support is too vast \parencite{Petersen_2012}. In the case of continuous exposures or mediators, we suggest checking positivity by categorizing the dependent variable into multiple classes and applying dePoRT to each, treating them as categorical. By default, PoRT uses the variable's quartiles, although we recommend prioritizing expert knowledge whenever possible. Recently, positivity diagnostic tools were proposed for a continuous exposure by \textcite{Moodie_Schulz_2025}, in regression settings, and by \textcite{Ring_Schomaker_2026} using kernel-based sparsity. However, their interpretability is challenging, and their use in mediation analysis needs further evaluation. Continuous confounders could also be slightly categorized into meaningful cut-offs to reduce the computational time of PoRT \parencite{Danelian_2023}.

There are several ways to address positivity violations. The most extreme is to target another estimand that is identifiable with the data at hand. While this shifts the scientific question one aims to answer, it is the only solution that works for a structural violation. A second solution is to select the decomposition that achieves the best positivity, as illustrated in our second application. Additionally, regression or imputation-based estimators may be favored over weighting estimators for practical violations if we allow for extrapolation \parencite[e.g.,][]{Lange_2012, Posch_2021, Imai_2010, Samoilenko_Lefebvre_2023}. Finally, one might balance the risks of violating positivity against those of residual confounding by excluding weak confounders — that is, variables expected to be strongly associated with the exposure but only weakly related to the outcome \parencite{Cole_Hernan_2008} or the mediator. This is especially likely to occur in high-dimensional settings.

Nonetheless, dePoRT is not a silver bullet. Its proper use presupposes an accurate \textit{a priori} identification of confounders, the correct temporal ordering of all variables, and the absence of measurement errors or misclassification;  evaluation of these relies heavily on domain expertise. In addition, dePoRT’s performance is influenced by PoRT's hyperparameters, $\alpha$, $\beta$, and $\gamma$ (and, to a lesser extent, those of the regression or classification tree). We recommend exploring a grid of pragmatically chosen values for them (see the Discussion in \textcite{Danelian_2023}).

\section{Conclusion}

This study extends the analysts' toolbox in causal mediation analysis for checking assumptions to achieve more robust results and transparent analyses. The PoRT framework is available for cross-sectional \parencite{Danelian_2023}, longitudinal \parencite{Chatton_2024}, and now mediation settings. It does not rely on parametric assumptions and can be used before applying any causal estimator. To encourage its broader adoption by the applied community, we have made dePoRT easy and free to use through the \texttt{R} package \texttt{port} \parencite{port_package} and a notebook available at github.com/ArthurChatton/dePoRT-notebook.

\printbibliography

\newpage
\onecolumn
\appendix

\section{Proofs} \label{proof}

The three proofs presented below are taken from \textcite{Nguyen_2022} and \textcite{VanderWeele_Vansteelandt_Robins_2014}.

\subsection{Controlled direct effect}

Assume the following identifiability assumptions:
\begin{assump} \label{cde_c}
   Consistency: $Y=Y^{a,m}$ if $A=a$ and $M=m$. 
\end{assump}
\begin{assump} \label{cde_ea}
   Exchangeability of the exposure: $A \ind Y^{a,m}|C$. 
\end{assump}
\begin{assump} \label{cde_em}
   Exchangeability of the mediator: $M \ind Y^{a,m}|A=a,C,L$.
\end{assump}
\begin{assump} \label{cde_pa}
   Positivity of the exposure: $\mathrm{P}(A=a|C=c) > \beta \geq 0, \forall c$ such as $\mathrm{P}(C=c)>0$.
\end{assump}
\begin{assump} \label{cde_pm}
   Positivity of the mediator: $\mathrm{P}(M=m|A=a,C=c,L=l) > \beta \geq 0, \forall l,c$ such as $\mathrm{P}(L=l,C=c)>0$. 
\end{assump}

\begin{proof}
\begin{align*}
   \mathrm{E}(Y^{a,m}) &=\sum_c\mathrm{E}(Y^{a,m}|C)\mathrm{P}(C=c) && \text{(iterated expectation)} \\
   & = \sum_c\mathrm{E}(Y^{a,m}|C=c,A=a)\mathrm{P}(C=c) && \text{\eqref{cde_ea} and \eqref{cde_pa}}\\
   & = \sum_{c,l}\mathrm{E}(Y^{a,m}|C=c,A=a,L=l)\mathrm{P}(L=l|C=c,A=a)\mathrm{P}(C=c) && \text{(iterated expectation)}\\
   & = \sum_{c,l}\mathrm{E}(Y^{a,m}|C=c,A=a,L=l,M=m)\mathrm{P}(L=l|C=c,A=a)\mathrm{P}(C=c) && \text{\eqref{cde_em} and \eqref{cde_pm}}\\
   & = \sum_{c,l}\mathrm{E}(Y|C=c,A=a,L=l,M=m)\mathrm{P}(L=l|C=c,A=a)\mathrm{P}(C=c). && \text{\eqref{cde_c}}
\end{align*}
\end{proof}

Note that our application in the main text did not consider $L$. Therefore, the final line of the proof simplifies to $\mathrm{E}(Y^{a,m}) = \sum_c\mathrm{E}(Y|C=c,A=a,M=m)\mathrm{P}(C=c)$.

\subsection{Natural effects}

Assume the following identifiability assumptions for the nested potential outcome's expectation $\mathrm{E}(Y^{a,M^{a^*}})$ (where $a^*$ can equal $a$):
\begin{assump} \label{ne_cy}
   Consistency of the nested potential outcome: $Y^{a,M^{a^*}}=Y^{a,m}$ if $M^{a^*}=m$.
\end{assump}
\begin{assump} \label{ne_cm}
   Consistency of the mediator: $M=M^{a}$ if $A=a$.
\end{assump}
\begin{assump} \label{ne_ea}
   Exchangeability of the exposure: $A \ind Y^{a,m}|C$ and $A \ind M^{a^*}|C$.
\end{assump}
\begin{assump} \label{ne_em}
   Exchangeability of the mediator: $M \ind Y^{a,m}|A=a,C$ and $M^{a^*} \ind Y^{a,m}|C$.
\end{assump}

\begin{assump} \label{ne_pm}
   Positivity of the mediator: $P(M=m|A=a,C=c) > \beta \geq 0, \forall c$ such as $P(M=m|A=a^*,C=c)>0$ and $\mathrm{P}(C=c)>0$. 
\end{assump}

\begin{proof}
\begin{align*}
   \mathrm{E}(Y^{a,M^{a^*}}) &=\sum_{c,m}\mathrm{E}(Y^{a,M^{a^*}}|M^{a^*}=m,C=c)\mathrm{P}(M^{a^*}=m|C=c)\mathrm{P}(C=c) && \text{(iterated expectation)} \\
    & = \sum_{c,m}\mathrm{E}(Y^{a,m}|M^{a^*}=m,C=c)\mathrm{P}(M^{a^*}=m|C=c)\mathrm{P}(C=c) && \text{\eqref{ne_cy}}\\
   & = \sum_{c,m}\mathrm{E}(Y^{a,m}|C=c)\mathrm{P}(M^{a^*}=m|C=c)\mathrm{P}(C=c) && \text{\eqref{ne_em} and \eqref{ne_pm}}\\
   & = \sum_{c,m}\mathrm{E}(Y^{a,m}|A=a,C=c)\mathrm{P}(M^{a^*}=m|C=c)\mathrm{P}(C=c) && \text{\eqref{ne_ea} and \eqref{cde_pa}}\\
    & = \sum_{c,m}\mathrm{E}(Y^{a,m}|A=a,M=m,C=c)\mathrm{P}(M^{a^*}=m|C=c)\mathrm{P}(C=c) && \text{\eqref{ne_em} and \eqref{ne_pm}}\\
     & = \sum_{c,m}\mathrm{E}(Y|A=a,M=m,C=c)\mathrm{P}(M^{a^*}=m|C=c)\mathrm{P}(C=c) && \text{\eqref{cde_c}} \\
     & = \sum_{c,m}\mathrm{E}(Y|A=a,M=m,C=c)\mathrm{P}(M^{a^*}=m|A=a^*,C=c)\mathrm{P}(C=c) && \text{\eqref{ne_ea} and \eqref{cde_pa}} \\
     & = \sum_{c,m}\mathrm{E}(Y|A=a,M=m,C=c)\mathrm{P}(M=m|A=a^*,C=c)\mathrm{P}(C=c). && \text{\eqref{ne_cm}} 
\end{align*}
\end{proof}

\subsection{Interventional effects}

The counterfactual random variable $\mathcal{M}^{a|C}$ presents the following two properties \parencite{VanderWeele_Vansteelandt_Robins_2014}:

\begin{prop}\label{ie_prop}
$\mathcal{M}^{a|C}$ follows the same conditional distribution given $C$ as $M^a$.
\end{prop}

\begin{prop}\label{ie_prop2}
$\mathcal{M}^{a|C}$ is randomized within strata of $C$ according to the distribution of $M^a$ such that it is conditionally independent of all other observed and counterfactual (nested or not) variables given $C$.
\end{prop}

Assume also the following identifiability assumptions for the nested potential outcome $Y^{a,\mathcal{M}^{a^*|C}}$ (where $a^*$ can equal $a$):
Consistency of the outcome \eqref{cde_c} and the mediator \eqref{ne_cm},
Exchangeability \eqref{ne_ea}, and positivity of the exposure \eqref{cde_pa}, and:

\begin{assump} \label{ie_em}
   Exchangeability of the mediator: $M \ind Y^{a,m}|A=a,C,L$.
\end{assump}
\begin{assump} \label{ie_pm}
   Positivity of the mediator: $P(M=m|A=a,C=c, L=l) > \beta \geq 0, \forall c$ such as $P(M=m|A=a^*,C=c)>0$ and $\mathrm{P}(C=c)>0$. 
\end{assump}

\begin{proof}
To reduce the notational burden, we reduce the random variables to their realization, except for observed and counterfactual mediators below.
\begin{align*}
   \mathrm{E}(Y^{a,\mathcal{M}^{a^*|C}}) &= \sum_{c,m}\mathrm{E}(Y^{a,\mathcal{M}^{a^*|C}}|c, \mathcal{M}^{a^*|C}=m)\mathrm{P}(\mathcal{M}^{a^*|C}=m|c)\mathrm{P}(c) && \text{(iterated expectation)}
 \\
 &= \sum_{c,m}\mathrm{E}(Y^{a,m}|c)\mathrm{P}(M^{a^*}=m|c)\mathrm{P}(c) && \text{\eqref{ie_prop} and \eqref{ie_prop2}} \\
 &= \sum_{c,m}\mathrm{E}(Y^{a,m}|c, a)\mathrm{P}(M^{a^*}=m|c, a^*)\mathrm{P}(c) && \text{\eqref{ne_ea} and \eqref{cde_pa}} \\
 &= \sum_{c,l,m}\mathrm{E}(Y^{a,m}|c, a, l)\mathrm{P}(l|c,a)\mathrm{P}(M^{a^*}=m|c, a^*)\mathrm{P}(c) && \text{(iterated expectation)} \\
 &= \sum_{c,l,m}\mathrm{E}(Y^{a,m}|c, a, l, M=m)\mathrm{P}(l|c,a)\mathrm{P}(M^{a^*}=m|c, a^*)\mathrm{P}(c) && \text{\eqref{ie_em} and \eqref{ie_pm}} \\
 &= \sum_{c,l,m}\mathrm{E}(Y|c, a, l, M=m)\mathrm{P}(l|c,a)\mathrm{P}(M=m|c, a^*)\mathrm{P}(c). && \text{\eqref{cde_c} and 
 \eqref{ne_cm}}
\end{align*}
\end{proof}

\section{Descriptive statistics for the application} \label{app:desc}

\begin{table*}[ht!]
\begin{center}
\caption{Characteristics of the included individuals from the International Dating Violence Study, 2001-2006.}
\begin{threeparttable}
\centering
\begin{tabular}{lccc}
  \hline
 & Overall & No CVE & CVE \\ 
& n=6224 & n=5049 & n=1175 \\ 
\hline
  Age, mean (SD) & 23.90 (7.05) & 23.88 (7.06) & 23.97 (7.01) \\ 
  Years in school, mean (SD) & 14.42 (1.17) & 14.42 (1.17) & 14.42 (1.21) \\ 
  FEZ, mean (SD) & -0.02 (0.99) & 0.00 (0.98) & -0.09 (1.02) \\ 
  MEZ, mean (SD) & -0.01 (1.00) & 0.01 (0.99) & -0.09 (1.01) \\ 
  FIZ, mean (SD) & 0.01 (1.00) & 0.04 (0.99) & -0.07 (1.02) \\ 
  PTSS, mean (SD) & 2.15 (0.54) & 2.10 (0.53) & 2.37 (0.52) \\ 
  Female, n (\%) & 4760 (76.5) & 3918 (77.6) & 842 (71.7) \\ 
  Married parents, n (\%) & 4143 (66.6) & 3446 (68.3) & 697 (59.3) \\ 
  Mixed sex couple, n (\%) & 6084 (97.8) & 4940 (97.8) & 1144 (97.4) \\ 
  Religious involvement &  &  &  \\ 
  \quad High, n (\%)  & 1143 (18.4) & 882 (17.5) & 261 (22.2) \\
  \quad Moderate, n (\%) & 616 (9.9) & 469 (9.3) & 147 (12.5) \\
  \quad Low, n (\%) & 2940 (47.2) & 2393 (47.4) & 547 (46.6) \\
  \quad None, n (\%) & 1525 (24.5) & 1305 (25.8) & 220 (18.7) \\
   \hline
\end{tabular}
\vspace{0.1cm}
\begin{tablenotes}
\item Abbreviations: CVE, Childhood exposure to violence; FEZ, Father education z-score; FIZ, Family income z-score; MEZ, Mother education z-score; PTSS, Post-traumatic stress symptoms. 
\end{tablenotes}
\end{threeparttable}
\end{center}
\end{table*}

\newpage
\section{Mediator positivity for natural effects} \label{app:decomp}

\subsection{Decompositions into direct and indirect effects}

Two decompositions coexist and are equally valid for the natural effects. The mediator positivity assumption differs between them. As such, when there is no substantive reason to prefer one over the other, the verification of their positivity assumptions could be used to choose the most suitable decomposition according the data one has at hand.

Recall the g-formula for the total effect: $\mathrm{E}(Y^a)= \sum_c\mathrm{E}(Y|A=a, C=c)\mathrm{P}(C=c)$. Consider a specific subgroup $c'$ (e.g., men younger than 20 years) for which no individual is exposed ($\mathrm{P}(A=1|C=c')=0$). It follows that one cannot compute the outcome conditional expectation ($\mathrm{E}(Y|A=1,C=c')$) and this quantity is thus undefined. In that case, the expectation of the potential outcome is also undefined, except when the subgroup $c'$ cannot be observed (i.e., $\mathrm{P}(C=c')=0$) because the product of the terms in the mediational formula would then be zero. Note that when continuous variables are present in $C$, the sum is replaced by an integral.

We can apply the same reasoning to the natural effects using the mediational formula \eqref{medform}. We have \begin{equation}
    \label{y1m0} \mathrm{E}(Y^{1,M^{0}})=  \sum_m\sum_c\mathrm{E}(Y|A=1, M=m, C=c)  \mathrm{P}(M=m|A=0,C=c)\mathrm{P}(C=c) \end{equation} and \begin{equation}\label{y0m1} \mathrm{E}(Y^{0,M^{1}})=  \sum_m\sum_c\mathrm{E}(Y|A=0, M=m, C=c)  \mathrm{P}(M=m|A=1,C=c)\mathrm{P}(C=c). \end{equation} When, in an observed specific subgroup $c'$, there are no exposed individuals (i.e., $\mathrm{P}(A=1|C=c')=0, \text{ and }\mathrm{P}(C=c')>0$), the quantity $\mathrm{E}(Y|A=1,M=m, C=c')$ cannot be directly estimated, and the potential outcome $Y^{1,M^{0}}$ is undefined. Thus, one also needs the exposure-related part of the positivity assumption for natural effects estimation. 
The mediator-related part of the positivity assumption can be understood in a similar manner. 
First, consider the case when there are no individuals with a certain value of the mediator $M=m'$, regardless of the exposure value (i.e., $\mathrm{P}(M=m'|A=1,C=c')=0$ and $\mathrm{P}(M=m'|A=0,C=c')=0$). In that case, the outcome conditional expectations in both \eqref{y1m0} and \eqref{y0m1} are not directly estimable. However, since these expectations are multiplied by the conditional probability of the mediator in the other exposure group, which is equal to zero (e.g., $\mathrm{E}(Y|A=1, M=m', C=c')$ is multiplied by $\mathrm{P}(M=m'|A=0,C=c) = 0$ in \eqref{y1m0}), the potential outcomes' expectations are still identifiable. In contrast, when there are individuals for whom the mediator's value $M=m'$ is observed in one exposure group but not in the other (e.g., $\mathrm{P}(M=m'|A=1,C=c')=0$ \textit{but} $\mathrm{P}(M=m'|A=0,C=c')>0$), the story is different. In this example, while the product $\mathrm{E}(Y|A=0, M=m', C=c') \times \mathrm{P}(M=m'|A=1,C=c)$ is still equal to zero, the other product $\mathrm{E}(Y|A=1, M=m', C=c') \times \mathrm{P}(M=m'|A=0,C=c)$ is no longer defined. Thus,  $\mathrm{E}(Y^{0,M^{1}})$ is identifiable, while $\mathrm{E}(Y^{1,M^{0}})$ is not.

\subsection{Toy simulation}
To provide a deeper understanding of the differential bias between the natural effect decompositions, consider the following toy simulation study. We simulated three independent confounders, the exposure, the two potential mediators and the four potential outcomes using the following set of equations:
\begin{equation*} 
\begin{split}
C_1 &\sim Uniform(0,2), \text{ rounded at the nearest integer} \\
C_2 &\sim Bernoulli(0.5) \\
C_3 &\sim Bernoulli(0.5) \\
A &\sim Bernoulli(expit(-0.2 + 1.2 C_1 -0.4 C_2 - 0.4C_3)) \\
M^a &\sim Bernoulli(expit(-0.5 + 2 a + 0.2 C_1 + 0.5C_2 + 0.2 C_3 + 50a\mathds{1}(C_1=0))) \\
M &= AM^1 + (1-A)M^0\\
Y^{a,M^{a^*}} &\sim Bernoulli(expit(0.4 a + M^{a^*} - 0.8 C_1 + 0.8 C_2 + 0.35 C_3)), \text{ where $a^*$ can equal $a$} \\
Y &=AY^{1,M^1} + (1-A)Y^{0,M^0}
\end{split}
\end{equation*}

We added a mediator positivity violation on the potential mediator $M^a$ by including a strong interaction between $a$ and $\mathds{1}(C_1=0)$. On one hand, no individuals with $M=0$ are present among those with both $A=1$ and $C_1=0$. On the other hand, individuals with $A=0$ may have any value $m$. Thereby, a mediator positivity violation is expected for $Y^{1,M^0}$ but not $Y^{0,M^1}$. Note that there was no violation of the exposure positivity. We simulated 1,000, 5,000 and 10,000 individuals to avoid practical positivity violations. We used the weighted natural effect model of \textcite{Lange_2012} implemented in the \texttt{R} package \texttt{medflex} \parencite{medflex} to estimate the potential outcome's value. We performed 1,000 simulations, denoted $n_{sim}$. Our main performance measure is the relative bias, defined as $100\times n_{sim}^{-1}\left(\frac{\widehat{\mathrm{E}}(Y^{a,m})}{\mathrm{E}(Y^{a,m})} - 1\right)$. As expected, both $\hat{\mathrm{E}}(Y^{1})$ and $\hat{\mathrm{E}}(Y^{0})$ were unbiased and $\hat{\mathrm{E}}(Y^{1,M^0})$ was biased (Table \ref{tab:sim}). Surprisingly, $\hat{\mathrm{E}}(Y^{0,M^1})$ also presented a slight bias, likely attributable to a misspecification of the mediator model induced by the positivity violation simulated on $M$. In this toy example, the potential outcome $Y^{0,M^1}$ is identifiable, whereas $Y^{1,M^0}$ is not due to mediator positivity violation.

\begin{table}[h]
    \centering
    \small
    \caption{Relative bias (\%) for each potential outcome in the presence of a mediator positivity violation.}
    \label{tab:sim}
    \begin{tabular}{ccccc}
    \hline
       n  & $\mathrm{E}(Y^{1})$ & $\mathrm{E}(Y^{0}$) & $\mathrm{E}(Y^{1,M^0})$ & $\mathrm{E}(Y^{0,M^1})$\\
    \hline
       1,000  & -0.1 & 0.4 & -3.1 & -1.0 \\
       5,000  & 0.0 & -0.1 & -3.1 & -1.3 \\
      10,000 & -0.0 & 0.0 & -3.1 & -1.2 \\
    \hline
    \end{tabular}
\end{table}

\end{document}